\documentclass[conference]{IEEEtran}
\ifCLASSINFOpdf
\else
\fi
\usepackage{amssymb}
\usepackage{amsfonts}
\usepackage{amsmath}
\usepackage{graphicx}
\usepackage{multirow}
\usepackage{subcaption}
\usepackage{array}
\usepackage{cite}

\begin{document}
\title{A network approach for power grid robustness against cascading failures}
\author{\IEEEauthorblockN{Xiangrong Wang$ ^1 $, Yakup Ko\c c$ ^{1,2} $, Robert E. Kooij$ ^{1,3} $, Piet Van Mieghem$ ^1 $}
\IEEEauthorblockA{$ ^1 $Faculty of Electrical Engineering, Mathematics and Computer
Science, Delft University of Technology, The Netherlands\\
$ ^2 $Systems Engineering Section, Faculty of Technology, Policy and Management, Delft University of Technology, The Netherlands\\
$ ^3 $TNO (Netherlands Organization for Applied Science Research), Information and Communication Technology, The Netherlands
}
}


%Telephone: (800) 555--1212, Fax: (888) 555--1212}


\maketitle

%\begin{abstract}
%Cascading failures are one of the main reasons for blackouts in electrical power grids. Stable power supply requires a robust design of the power grid topology. Currently, the impact of the grid structure on the grid robustness is purely assessed by topological metrics, that fail to capture the fundamental properties of the electrical power grids such as power flow allocation according to Kirchhoff's laws. This paper proposes the usage of the effective graph resistance as a metric for the addition of single links to improve the grid robustness against cascading failures. Four strategies based on structural and spectral properties of a grid are investigated to improve the robustness at a lower computational complexity compared to that of exhaustive search. Exhaustively adding all the possible lines (one line at each time) occasionally shows a decrease of the grid robustness. We attribute this phenomenon to Braess's paradox that is originally found in traffic networks. We investigate the impact of the topology on Braess's paradox in power grids. Experimental verification shows that the effective graph resistance is an effective metric for identifying single links whose addition improves the robustness of power grids while avoids the Braess's paradox. Most of the strategies identify a line whose addition highly increases the grid robustness.
%\end{abstract}
\begin{abstract}
Cascading failures are one of the main reasons for blackouts in electrical power grids. Stable power supply requires a robust design of the power grid topology. Currently, the impact of the grid structure on the grid robustness is mainly assessed by purely topological metrics, that fail to capture the fundamental properties of the electrical power grids such as power flow allocation according to Kirchhoff's laws. This paper deploys the effective graph resistance as a metric to relate the topology of a grid to its robustness against cascading failures. Specifically, the effective graph resistance is deployed as a metric for network expansions (by means of transmission line additions) of an existing power grid. Four strategies based on network properties are investigated to optimize the effective graph resistance, accordingly to improve the robustness, of a given power grid at a low computational complexity. Experimental results suggest the existence of Braess's paradox in power grids: bringing an additional line into the system occasionally results in decrease of the grid robustness. This paper further investigates the impact of the topology on the Braess's paradox, and identifies specific sub-structures whose existence results in Braess's paradox. Careful assessment of the design and expansion choices of grid topologies incorporating the insights provided by this paper optimizes the robustness of a power grid, while avoiding the Braess's paradox in the system.
\end{abstract}


% \fi
\IEEEpeerreviewmaketitle

\section{Introduction}
The electrical power grid is crucial for economic prosperities of modern societies. Disruptions to electrical power grids paralyze the daily life and cause huge economical and social costs for these societies \cite{wang2011robustness, chang2011performance, kocc2013entropy}. The strong dependency of other crucial infrastructures such as telecommunication, transportation and water supply on electrical power grids amplifies the severity of large scale blackouts \cite{hines2009cascading}. The key importance of the power grid encourages further research into sustaining power system reliability and developing new approaches to evaluate and mitigate the risk of cascading blackouts. 

Cascading failures are one of the main reasons for large scale blackouts \cite{dobson2014cascading}. Cascading failures are the consequence of the collective dynamics of a complex power grid. Large scale cascades are typically due to the propagation of a local failure into the global network \cite{zhang2013understanding}. Consequently, analyzing and mitigating cascading failures requires a system level approach. Recent advances in the field of network science \cite{dong2013robustness,karrer2014percolation} provide the promising potential of complex network theory to investigate the robustness of power grids at a system level. The robustness of power grids in this paper refers to their maintenance of function after cascading failures triggered by targeted attacks.

Analyzing and improving the network robustness includes two parts. The first goal is the proposal of a proper metric that characterizes the robustness of a specific class of networks \cite{trajanovski2013robustness}. A second goal is to propose efficient strategies on graph modification in order to increase the value of the proposed robustness metric. Consequently, an effective robustness metric that incorporates the essence of the power grids and effective strategies for graph modification are required to improve the robustness of power grids.

The effective graph resistance is a graph metric which characterizes the essence of electrical power grids such as power flow allocation according to Kirchhoff's laws. Researchers in \cite{kocc2014impact} show that the effective graph resistance effectively captures the impact of cascading failures in a power grid. The lower the effective graph resistance is, the more robust a power grid is against cascading failures. Adding a link decreases the effective graph resistance \cite{Ellen2011Effective}. This paper focuses on enhancing the grid robustness against cascading failures by applying the effective graph resistance as a metric for network expansion. 
%. 

Determining the right pair of nodes to connect in order to maximize the robustness is a challenge. Exhaustive search, i.e. checking all the possibilities, is computationally expensive. Compared to exhaustive search, this paper proposes four strategies that provide a trade-off between a higher decrease of the effective graph resistance and a lower computational complexity. 

Exhaustively evaluating the impact of each link addition on robustness reveals the occurrence of Braess's paradox in power grids. Braess's paradox, originally found in traffic networks \cite{braess2005paradox}, shows that adding a link can decrease the robustness of the network. Specific sub-structures that might result in Braess's paradox by adding an extra link are investigated. Simulation results indicate that the effective graph resistance effectively identifies a link whose addition increases the robustness while avoids the Braess's paradox. Moreover, most of the strategies highly increase the robustness at a low computational complexity.

This paper is organized as follows: Section \ref{Section_ModelCascadingFailure} introduces the model of cascading failures in power grids. Section \ref{Section_EffectiveGraphResistance} presents the computation of the effective graph resistance in power grids. Strategies to add a transmission line are illustrated in Section \ref{Section_StrategiesAddingTransmissionLine}. The experimental methodology is illustrated in Section \ref{Section_ExperimentalMethodology} and the improvement of the grid robustness is evaluated in Section \ref{Section_NumericalAnalysis}. Section \ref{Section_Conclusion} concludes the paper. 

\section{Model of Cascading Failures in Power Grids}
\label{Section_ModelCascadingFailure}
A power grid is a three-layered network consisting of generation, transmission and distribution parts. A graph can represent a power grid where nodes are generation, transmission, distribution buses and substations, and links are transmission lines. Additionally, links are weighted by the admittance (or impedance) values of the corresponding transmission lines.
  
Electrical power in a grid is distributed according to Kirchoff's laws. Accordingly, impedances, voltage levels at each individual power station, voltage phase differences between power stations and loads at terminal stations control the power flow in the grid. This paper approximates the flow values in a grid by using a linear DC flow equation \cite{VanHertem2006AC} that approximates the nonlinear AC power flow equation \cite{grainger1994power}. 
   
The maximum capacity $ C_{l} $ of a line $ l $ is defined as the maximum power flow that can be afforded by the line. As in \cite{kocc2014impact}, we assume that the maximum capacity of a transmission line is proportional to its initial load $ L_l(0) $ as follows:
\begin{equation}
C_{l}=\alpha_lL_l(0)
\label{maxCapacity}
\end{equation}
where $ \alpha_l $ is called the tolerance parameter of the line $ l $.

In a power grid, transmission lines are protected by relays and circuit breakers. A relay of a transmission line measures the load of that line and compares the load with the maximum capacity $ C_{l} $ computed by equation (\ref{maxCapacity}). When the maximum capacity is violated, and this violation lasts long enough, the relay notifies a circuit breaker to trip the transmission line in order to prevent the line from permanent damage due to overloading. We assume a deterministic model for the line tripping mechanism. A circuit breaker trips at the moment the load of a transmission line exceeds its maximum capacity.

The failure of a transmission line changes the balance of the power flow distribution over the grid and causes a redistribution of the power flow over the network. This dynamic response of the system to this triggering event might overload other transmission lines in the network. The protection mechanism trips these newly overloaded transmission lines and the power flow is again redistributed potentially resulting in new overloads. This cascading failure continues until no more transmission lines are overloaded.
\section{Effective Graph Resistance in Power Grids}
\label{Section_EffectiveGraphResistance}
This section explains the complex network preliminaries, presents the effective graph resistance, and elaborates on how it is computed in electric power grids.
\subsection{Complex Network Preliminaries}
The topology of complex networks can be represented by a graph $ G(N,L) $ consisting of $ N $ nodes connected by $ L $ links. Graphs with $ N $ nodes are completely described by an $ N \times N $ adjacency matrix $ A $, in which the element $ a_{ij}=1 $ if there is a link between nodes $ i $ and $ j $, otherwise $ a_{ij}=0 $. In case of a weighted graph, the network is represented by the weighted adjacency matrix $ W $ where the element $ w_{ij} $ is a real number that characterizes a certain property of the link $ i \sim j $. The weight can be distances in transportation networks, the delay in the Internet, the strength of the interaction in the brain networks, and so on. 

The weighted Laplacian matrix $Q=\Delta-W$ of $G$ is an $ N \times N $ matrix, where $ \Delta=\text{diag}(d_i) $ is the $ N \times N $ diagonal degree matrix with the element $ d_i=\sum_{j=1}^{N}w_{ij} $. The eigenvalues of $Q$ are non-negative and at least one is zero \cite{PVM_graphspectra}. Thus, the smallest eigenvalue of $ Q $ is zero. The eigenvalues of $Q$ are ordered as $%
0=\mu _{N}\leq \mu _{N-1}\leq \ldots \leq \mu _{1}$. 

Graph metrics measure the structural and spectral properties of networks. The degree $ d_i $ of a node $ i $ specifies the number of connected neighbours to that node. The largest eigenvalue $ \lambda_1 $ (also called the spectral radius) of the adjacency matrix highly influences the dynamic processes on networks such as virus spreading and synchronization processes \cite{Restrepo2007Approximating}. The eigenvector corresponding to the spectral radius is called principle eigenvector $ x_1 $ that characterizes the influence of link/node removal on spectral radius \cite{van2011decreasing,li2012bounds}. The second smallest eigenvalue $ \mu _{N-1} $ of the Laplacian matrix $ Q $ is coined by Fiedler \cite{Fiedler} as the algebraic connectivity $ \alpha_G $. The corresponding eigenvector is called the Fiedler vector. The entries of the Fiedler vector provide a powerful heuristic for community detection and graph partitioning \cite{newman2013community}. The strategies illustrated in Section \ref{Section_StrategiesAddingTransmissionLine} are based on these structural and spectral graph metrics.

\subsection{Effective graph resistance in power grids}
Effective resistance $ R_{ij} $ is the electrical resistance between nodes $ i $ and $ j $ computed by series and parallel manipulations when a graph is seen as an electrical circuit where each link in the graph has a unit resistance. According to the Ohm's law, the effective resistance is the potential difference between nodes $ i $ and $ j $ when a unit current is injected at node $ i $ and withdrawn at node $ j $. The effective graph resistance $ R_G $ is the sum of the effective resistance over all pairs of nodes $ R_G=\sum_{i=1}^{N}\sum_{j=i+1}^{N}R_{ij} $. 
%Effective graph resistance takes into account all the paths between two nodes and the total weight on each path.

Computation of the effective graph resistance for a power grid necessitates the topology of the grid (i.e. interconnection of nodes) and reactance (or susceptance) values of the transmission lines in the grid. The weighted Laplacian matrix $ Q $ of a power grid reflects the interconnection of nodes by transmission lines. The weight $ w_{ij} $ corresponds to the susceptance (the inverse of reactance) value of the line $ l=i \sim j $. The effective resistance $ R_{ij} $ between a pair of nodes is computed as \cite{PVM_graphspectra}:
\begin{equation}
R_{ij}=\left(\hat{Q}^{-1}\right)_{ii}+\left(\hat{Q}^{-1}\right)_{jj}-2\left(\hat{Q}^{-1}\right)_{ij}
\label{PairwiseEffecResis}
\end{equation}
where $ \hat{Q}^{-1} $ is the Moore-Penrose pseudo-inverse of the $ Q $.

In terms of eigenvalues of the weighted Laplacian matrix $ Q $, the effective graph resistance can be written as \cite{PVM_graphspectra}
\begin{equation}
R_{G}=N\overset{N-1}{\underset{i=1}{\sum }}\frac{1}{\mu _{i}}
\label{EGR_Computation}
\end{equation}
where $ \mu_i $ is the $ i $th eigenvalue of $ Q $ and $ N $ is the number of nodes in a power grid. In this paper, we use equation \eqref{EGR_Computation}, which is computationally efficient, to compute the effective graph resistance.

\section{Strategies for Adding a Transmission Line}
\label{Section_StrategiesAddingTransmissionLine}

As a response to blackouts, additional transmission lines are placed aiming to increase the robustness of power grids. Determining the right pair of nodes to connect in order to maximize the robustness is the challenge. An exhaustive search, identifying the best pair of nodes to connect by checking all $ L_c=\binom{N}{2}-L $ possibilities, is computationally expensive especially when the number of nodes increases. Therefore, strategies
that determine the transmission line to be added based on topological and spectral properties of a network, provide a trade-off between a scalable computation and a high
increase of the grid robustness.

Topological and spectral metrics, such as degree, algebraic connectivity and spectral radius, characterize the connectivity of a network and highly influence the dynamic processes executed on a network \cite{Fiedler,van2012epidemic}. The effective graph resistance is shown to be able to anticipate the robustness of power grids with respect to cascading failures \cite{kocc2014impact}. This section investigates four strategies, studied in \cite{wang2014improving}, for selecting a link whose addition potentially minimizes the effective graph resistance and accordingly maximizes the robustness. A strategy defines a link $l=i\sim j$ and the selection of nodes $i$
and $j$ for each strategy are illustrated in the rest of this section.

\subsection{Degree product}

The nodes $i$ and $j$ have the minimum product of degrees $\min (d_{i}d_{j})$. If there are multiple node pairs with the same minimum product of degrees,
one of these pairs is randomly chosen.

The complexity for the strategy is $O(N^{2}-N+2L_{c})$ computed as
follows: (i) $O(N(N-1))$ is for counting the degrees of all the nodes. (ii) $%
O(2L_{c})$ is for computing $d_{i}d_{j}$ for $L_{c}$ unconnected node pairs and for finding $\min (d_{i}d_{j})$.

\subsection{Principle eigenvector}
The nodes $i$ and $j$ correspond to the $i^{th}$ and $%
j^{th}$ components of the principal eigenvector $x_{1}$ that have the
maximum product ${\max }((x_{1})_{i}(x_{1})_{j})$ of the principle eigenvector components. The principal eigenvector $x_{1}$ belongs to the largest eigenvalue of the
weighted adjacency matrix $ W $.

The complexity is $O(N^{3}+2L_{c})$ computed as
follows: (i) $O(N^{3})$ is for computing the principle eigenvector $x_{1}$ assuming
the adoption of the QR algorithm \cite{QRalgorithem} for computation. (ii) $%
O(2L_{c})$ is for computing $(x_{1})_{i}(x_{1})_{j}$ for $L_{c}$ unconnected node pairs and for finding $\max((x_{1})_{i}(x_{1})_{j})$.

\subsection{Fiedler vector}

The nodes $i$ and $j$ correspond to the $i^{th}$ and $j^{th}$ components of
the Fiedler vector $y$ that satisfy $\Delta y=\max (|y_{i}-y_{j}|)$, where $%
|y_{i}-y_{j}|$ is the absolute difference between the $i^{th}$ and $j^{th}$
components of the Fiedler vector \cite{HuijuanAlgebraicconnectivity}.

The complexity is $O(N^{3}+2L_{c})$ computed as
follows: (i) $O(N^{3})$ is for computing the Fiedler vector $y_{i}$ by the QR algorithm \cite{QRalgorithem}. (ii) $%
O(2L_{c})$ is for computing $|y_{i}-y_{j}|$ for $L_{c}$ unconnected node
pairs and for finding $\text{max}|y_{i}-y_{j}|$.

\subsection{Effective resistance}

The nodes $i$ and $j$ have the highest effective resistance $\max (R_{ij})$.
The pairwise effective resistance $R_{ij}$ is computed by equation \eqref{PairwiseEffecResis}. 

The complexity is $O(N^{3}+4L_{c})$ computed as
follows: (i) $O(N^{3})$ is for computing $\widehat{Q}^{-1}$. (ii) $O(4L_{c})$
is for computing $R_{ij}$ for $L_{c}$ unconnected node pairs and for finding the maximum $R_{ij}$.

Table \ref{Table_Strategy} summarizes all the strategies that identify a link $l=i\sim j$ and the order of their corresponding computational complexity. Table \ref{Table_Strategy} also presents the complexity order of the exhaustive search in order to compare with the complexity of the four strategies. The complexity order $ O(N^{5}) $ of the exhaustive search is computed by $ O(N^{2}) $ for checking all the possibilities multiplied by $ O(N^3) $ for computing the effective graph resistance after a link addition. 

\begin{table}
\centering\caption
{A summary of the strategies and the order of their computational complexity.}
\begin{tabular}{c|cc|c}
\hline
& Node $i$ & Node $j$ & Complexity Order \\ \hline

DegProd & \multicolumn{2}{|c|}{$\underset{i,j}{\arg \min }(d_{i}d_{j})$} & $%
O(N^{2})$ \\
PrinEigen & \multicolumn{2}{|c|}{$\underset{i,j}{\arg \max }\left(
(x_{1})_{i}(x_{1})_{j}\right)$}  & $%
O(N^{3})$ \\  
FiedlerVector & \multicolumn{2}{|c|}{$\underset{i,j}{\arg \max }(\left\vert
y_{i}-y_{j}\right\vert )$} & $O(N^{3})$ \\ 
EffecResis & \multicolumn{2}{|c|}{$\underset{i,j}{\arg \max }(R_{ij})$} & $%
O(N^{3})$ \\ 
Exhaustive Search & \multicolumn{2}{|c|}{$\underset{i,j}{\arg \min }(R_G)$} & $ O(N^5) $\\\hline
\end{tabular}
\label{Table_Strategy}
\end{table}
\section{Experimental Methodology}
\label{Section_ExperimentalMethodology}
The experimental method presented in this section evaluates the robustness of the improved power system against cascading failures triggered by deliberate attacks. This approach can be used to assess the performance of the effective graph resistance as a metric for link addition on improving the robustness of power grids. This section elaborates on attack strategies and the quantification of the grid robustness after cascading failures.
\subsection{Attack Strategies}
This paper designs attack strategies based on electrical node significance centrality and link betweenness centrality. The electrical node significance \cite{kocc2013entropy} is a flow-based measure for node centrality, specifically designed for power grids. The electrical node significance $ \delta_i $ of a node $ i $ is defined as the total power $ P_i $ distributed by node $ i $ normalized by the total amount of power that is distributed in the entire grid:
\begin{equation}
\delta_i=\frac{P_i}{\sum_{j=1}^{N}P_j}
\label{nodeSignificance}
\end{equation}
An attack based on $ \delta_i $ refers to target the link incident to the node $ i $ that has the highest electrical node significance. Since node $ i $ has the number $ d_i $ of incident links, the link with the highest load is chosen.

The link betweenness centrality is a topological graph metric quantifying the centrality of a link in complex networks \cite{PietPerformanceAnalysis}. The betweenness centrality of a link is defined as the total number of the shortest paths that traverse the link $ l $.
\begin{equation}
B_l=\sum_{i=1}^{N}\sum_{j=1}^{N} \text{1}_{l \in \mathcal{P}(i,j)}
\end{equation}
where $ \text{1}_{\{x\}} $ is the indicator function: $ \text{1}_{\{x\}}=1 $ if the condition $ {\{x\}} $ is true, else $ \text{1}_{\{x\}}=0 $, and $ \mathcal{P}(i,j) $ is the shortest path between nodes $ i $ and $ j $. An attack based on betweenness centrality targets the link with the highest betweenness centrality.   

Placing an additional line according to different strategies (presented in Section \ref{Section_StrategiesAddingTransmissionLine}) results in different improved power systems. In order to compare cascading damages of these improved systems, we always attack the same link identified by the node significance centrality or link betweenness centrality of the original power grid.
\subsection{Robustness Evaluation}
The robustness of power grids is evaluated by the criticality of the additional line and the damages after cascading failures triggered by targeted attacks. To assess the criticality of the newly added transmission line based on the effective graph resistance, we deploy an analogous approach as in \cite{latora2005vulnerability}: the criticality of an added line $ l $ in a graph $ G $ is determined by the relative decrease of the effective graph resistance $ \Delta R_G^l $ that is caused by the addition of a link $ l $:
\begin{equation}
\Delta R_G^l=\frac{R_G-R_{G+l}}{R_G}
\label{RobustnessEvaluation_EffGraphReis}
\end{equation}
where $ R_{G+l} $ is the effective graph resistance of the grid after adding a link $ l $ into $ G $. Evaluation of equation \eqref{RobustnessEvaluation_EffGraphReis} results in the theoretical robustness level of a power grid.

Initially, a transmission line identified by the four strategies and exhaustive search is added into the power grid. Then, the newly obtained grids are attacked and the cascading damages are quantified.

The damage caused by the cascade is quantified in terms of normalized served power demand $ DS $: served power demand divided by the total power demand in the network. Computing the normalized served demand for an interval of tolerance parameters $ [\alpha_{min}, \alpha_{max}] $ results in a robustness curve of a grid. The normalized area below the robustness curve is computed by a Riemann sum \cite{PietPerformanceAnalysis}:
\begin{equation}
r=\frac{\sum_{i=1}^{m+1}DS(\alpha_i)\Delta_{\alpha}}{\alpha_{max}-\alpha_{min}}
\label{SimulationEvaluation}
\end{equation}
where the closed interval $ [\alpha_{min}, \alpha_{max}] $ is equally partitioned by $ m $ points and the length of the resulting interval is $ \Delta_{\alpha}= \frac{\alpha_{max}-\alpha_{min}}{m+1} $. $DS(\alpha_i) $ is the normalized served demand when the tolerance parameter of the network is $ \alpha_i \in [\alpha_{min}+(i-1)\Delta_{\alpha},\alpha_{min}+i\Delta_{\alpha}]$. Since the maximum value of DS is $ 1 $, $ (\alpha_{max}-\alpha_{min} )$ refers to the maximum possible area below the robustness curve ensuring that the value of $ r $ is between $ 0 $ and $ 1 $. Evaluation of equation \eqref{SimulationEvaluation} for the robustness curve results in the experimental robustness level of a power grid with respect to cascading failures.
\section{Numerical Analysis}
\label{Section_NumericalAnalysis}
This section investigates the effectiveness of the effective graph resistance as a metric for line addition, the impact of structures on the Braess's paradox, and the performance of the four strategies. First, the power grid is expanded by adding single links according to the minimization of the effective graph resistance, and the criteria of the four strategies. Then, the robustness of the improved power grid is assessed quantitatively under targeted attacks. 
\subsection{Assessing effectiveness of the effective graph resistance}
Exhaustively adding all the possible links provides us all the possibly improved grids. Quantifying the cascading damages of all the improved grids under targeted attacks provides the benchmark for the evaluation of the effective graph resistance. The reactance value on each added line is assumed to be the average of all the existing transmission lines. The simulations are performed by MATCASC \cite{kocc2013matcasc}, a MATLAB based cascading failures analysis tool implementing the model in Section \ref{Section_ModelCascadingFailure}. 

\begin{figure}[!htp]\centering
\begin{subfigure}[b]{0.491\columnwidth}
\includegraphics[width=\columnwidth]{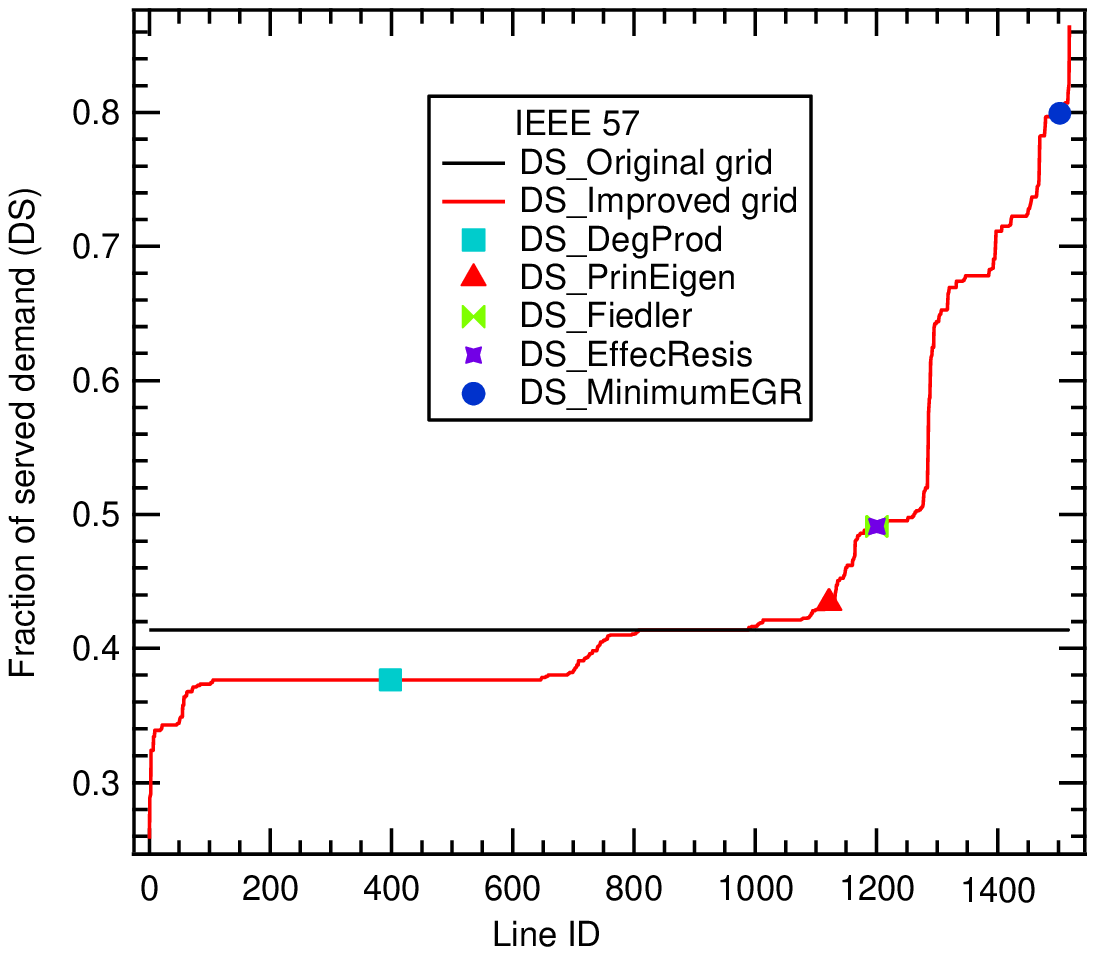}
\caption{IEEE 57}
\end{subfigure}
\begin{subfigure}[b]{0.491\columnwidth}
\includegraphics[width=\columnwidth]{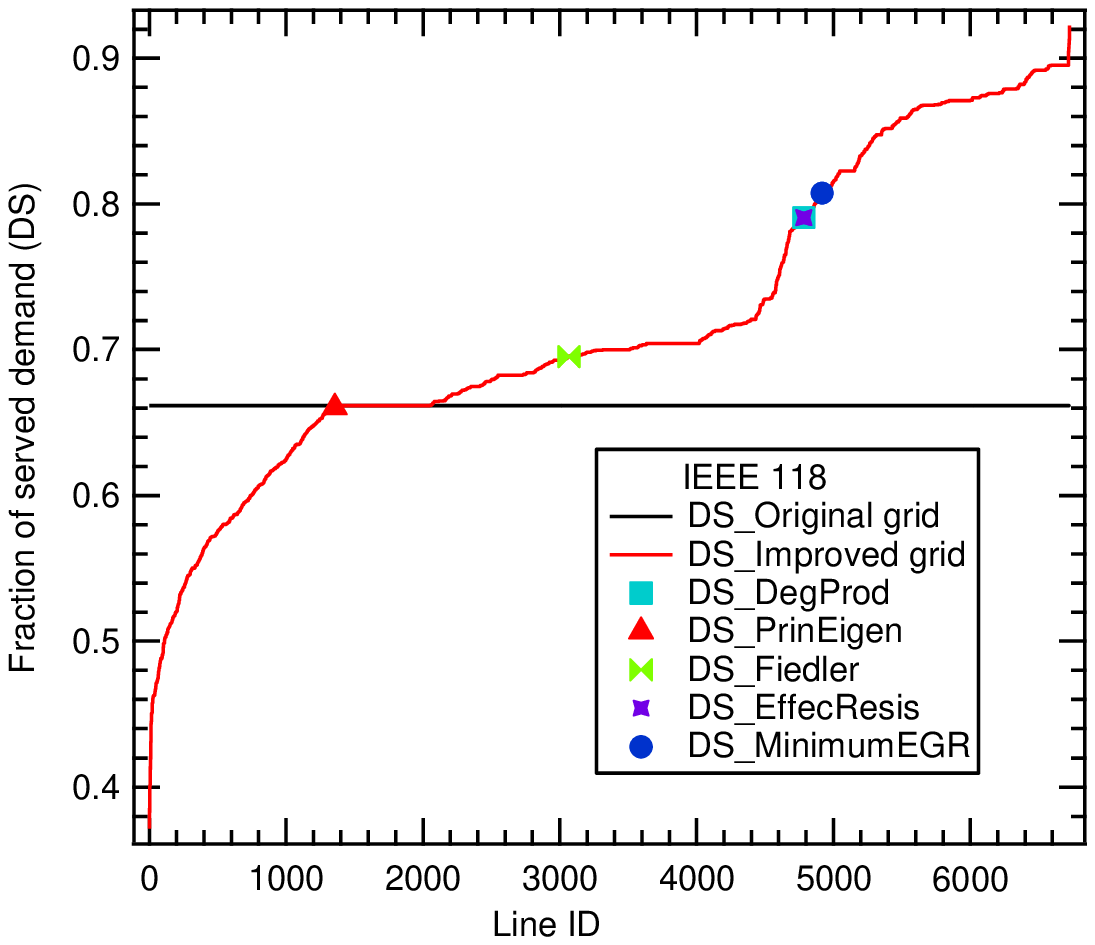}
\caption{IEEE 118}
\end{subfigure}
\caption{The performance of the effective graph resistance in IEEE 57, IEEE 118 power system with the tolerance parameter $ \alpha=2 $.}
\label{Tolerance_DS_ExhaustiveSearch}
\end{figure}

Figure \ref{Tolerance_DS_ExhaustiveSearch} shows the performance of the effective graph resistance on identifying a critical link under a fixed tolerance parameter $ \alpha=2 $ in IEEE 57 and 118 power systems. There are $ 1518 $ possible improved grids by adding a line to IEEE 57 and $ 6724 $ possible improved grids to IEEE 118. The original and improved power systems are attacked based on the node significance centrality computed by equation \eqref{nodeSignificance}. In Figure \ref{Tolerance_DS_ExhaustiveSearch}, the horizontal line (i.e. the black line) is the served demand $ DS $ for the original power grid after cascading failures. The points on the red curve refer to the DS value of each improved grid that is obtained by adding one single line to the original network. 

%Braess's paradox generally occurs in most complex networks, such as oscillator networks \cite{witthaut2012braess}, mechanical and electrical networks \cite{cohen1991paradoxical}, hydraulic networks and other networks that obey Kirchhoff's laws \cite{penchina2003braess}. In particular, Braess's paradox also exists in power systems \cite{blumsack2007quantitative,witthaut2012braess}. Our results are in line with the occurrence of Braess's paradox in power systems.

The performance of the effective graph resistance as a metric for link addition and the performance of strategies are labelled in the Figure \ref{Tolerance_DS_ExhaustiveSearch} with markers. The added line that minimizes the effective graph resistance increases the robustness from $ 0.41 $ to $ 0.80 $ and improves the robustness by $ 95\% $. Compared to the possibly maximal increase $ 0.86 $ by a single link addition, the effective graph resistance achieves $ 93\% $ accuracy in the IEEE 57 power system. Similarly in IEEE 118 power system, the added line that minimizes the effective graph resistance increases the robustness from $ 0.66 $ to $ 0.81 $. The effective graph resistance achieves $ 87 \% $ accuracy identifying the optimal line in the IEEE 118 power system.

In Figure \ref{Tolerance_DS_ExhaustiveSearch}, the curve above the horizontal line shows an increase of the robustness after a link addition, while the curve below the horizontal line presents a decrease of the robustness by adding a link. This counter-intuitive phenomenon is linked to Braess's paradox known for traffic networks, stating that adding extra capacity or links to a network occasionally reduces the overall performance of a network \cite{braess2005paradox}.  

The simulation results in Figure \ref{Tolerance_DS_ExhaustiveSearch} illustrate the effectiveness of the effective graph resistance to identify a critical link. The addition of the critical link improves the robustness of power grids regardless of the fact that the robustness can be decreased according to Braess's paradox. We further investigate more details on Braess's paradox in subsection \ref{subsection_BraessParadox}.
\subsection{Assessing the effectiveness of strategies}
To assess the effectiveness of the four strategies in Section \ref{Section_StrategiesAddingTransmissionLine}, the IEEE 118 power system, consisting of $ 118 $ buses and $ 186 $ lines, is considered as a use case. For each line identified by each strategy, equation \eqref{RobustnessEvaluation_EffGraphReis} is evaluated and its impact on the effective graph resistance is determined. Table \ref{CriticalAddedLine} shows the lines to be added identified by strategies and their impact on the decrease of $ R_G $.
\begin{table}[!htp]\centering
\renewcommand{\arraystretch}{1.5}
\begin{tabular*}{200pt}{@{\extracolsep{\fill}}ccc}
\hline Strategy & line ID & $ \Delta R_{G}^l $(\%) \\ 
\hline DegProd  & $ l_{87-117} $ & 9.0 \\ 
  PrinEigen & $ l_{87-111} $ & 4.2 \\ 
  Fiedler & $ l_{111-117} $ & 11.3 \\ 
  EffectiveResis & $ l_{87-117} $ & 9.0 \\ 
%  Exhaustive Search & $ l_{12-105} $ & 14.9 \\
\hline
\end{tabular*} 
\caption{Added lines identified by the strategies 
%and exhaustive search 
and their impact on the decrease of $ R_G $.} 
\label{CriticalAddedLine}
\end{table}

%Table \ref{CriticalAddedLine} illustrates that adding the line connecting bus $ 12 $ and bus $ 105 $ results in $ 14.9\% $ decrease of the effective graph resistance. 
In Table \ref{CriticalAddedLine}, the strategy based on the Fiedler vector selects the line connecting bus $ 111 $ and bus $ 117 $ and its addition causes $ 11.3\% $ decrease of the effective graph resistance. Strategies based on the degree product and the effective resistance have an equal performance that decrease the effective graph resistance by $ 9\% $. The strategy based on the principle eigenvector decreases the effective graph resistance by $ 4.2\% $. Compared to other strategies, the strategy based on the Fiedler vector performs the best.

\begin{figure}[!htp]\centering
\begin{subfigure}[b]{0.491\columnwidth}
\includegraphics[width=\columnwidth]{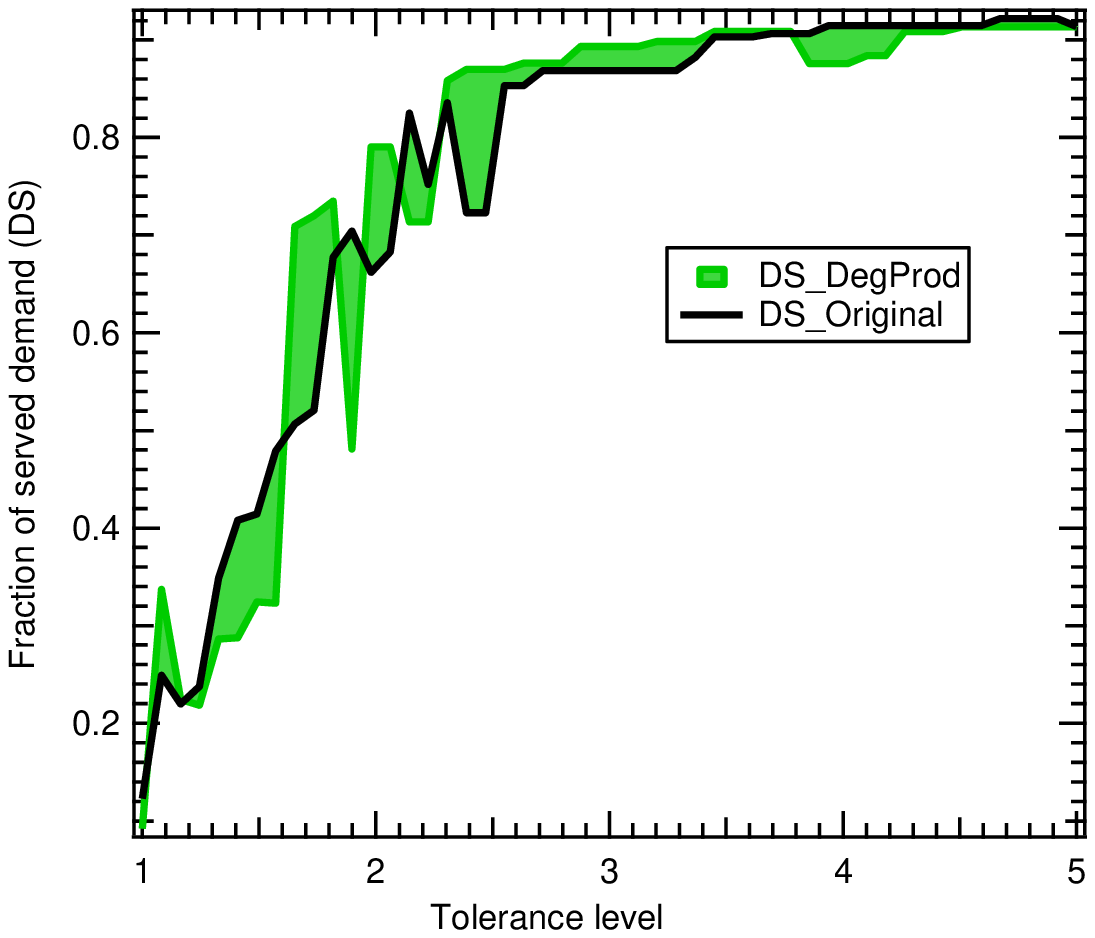}
\end{subfigure}
\begin{subfigure}[b]{0.491\columnwidth}
\includegraphics[width=\columnwidth]{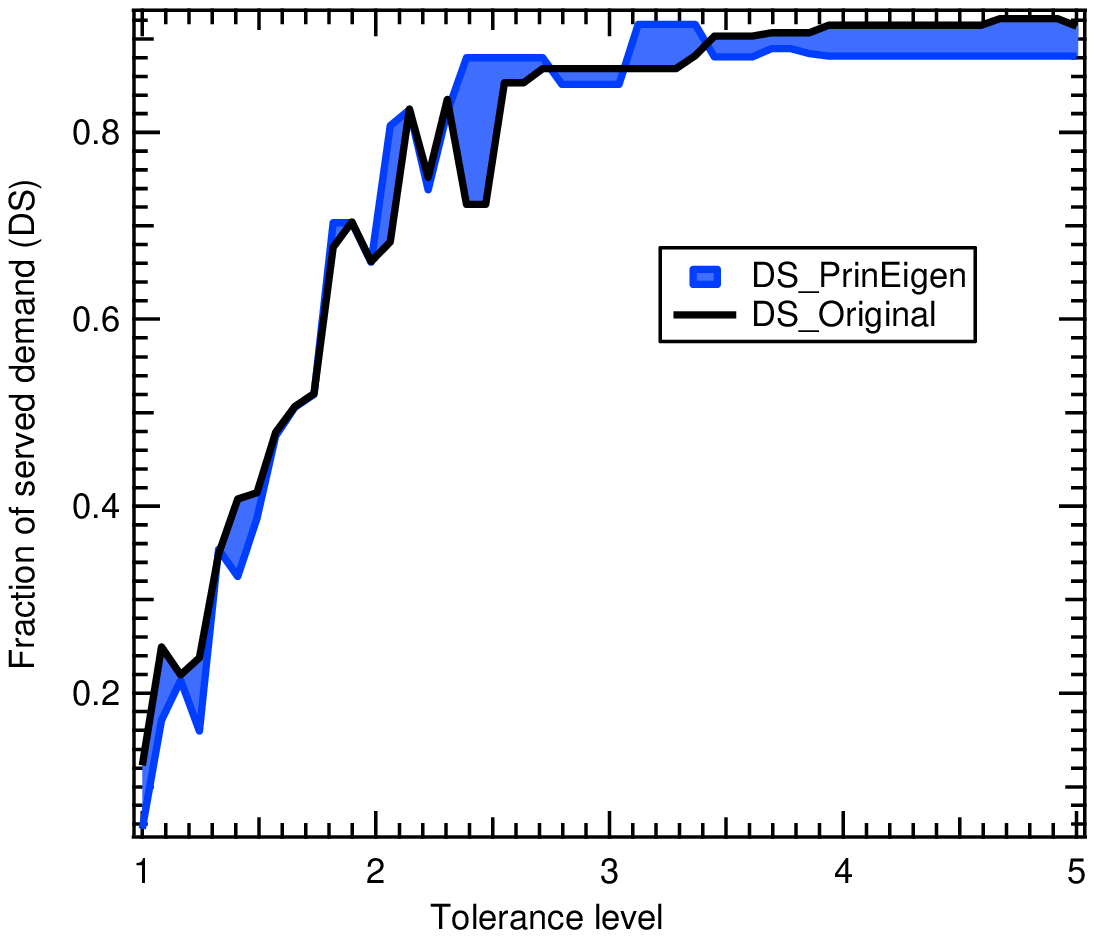}
\end{subfigure}
\begin{subfigure}[b]{0.491\columnwidth}
\includegraphics[width=\columnwidth]{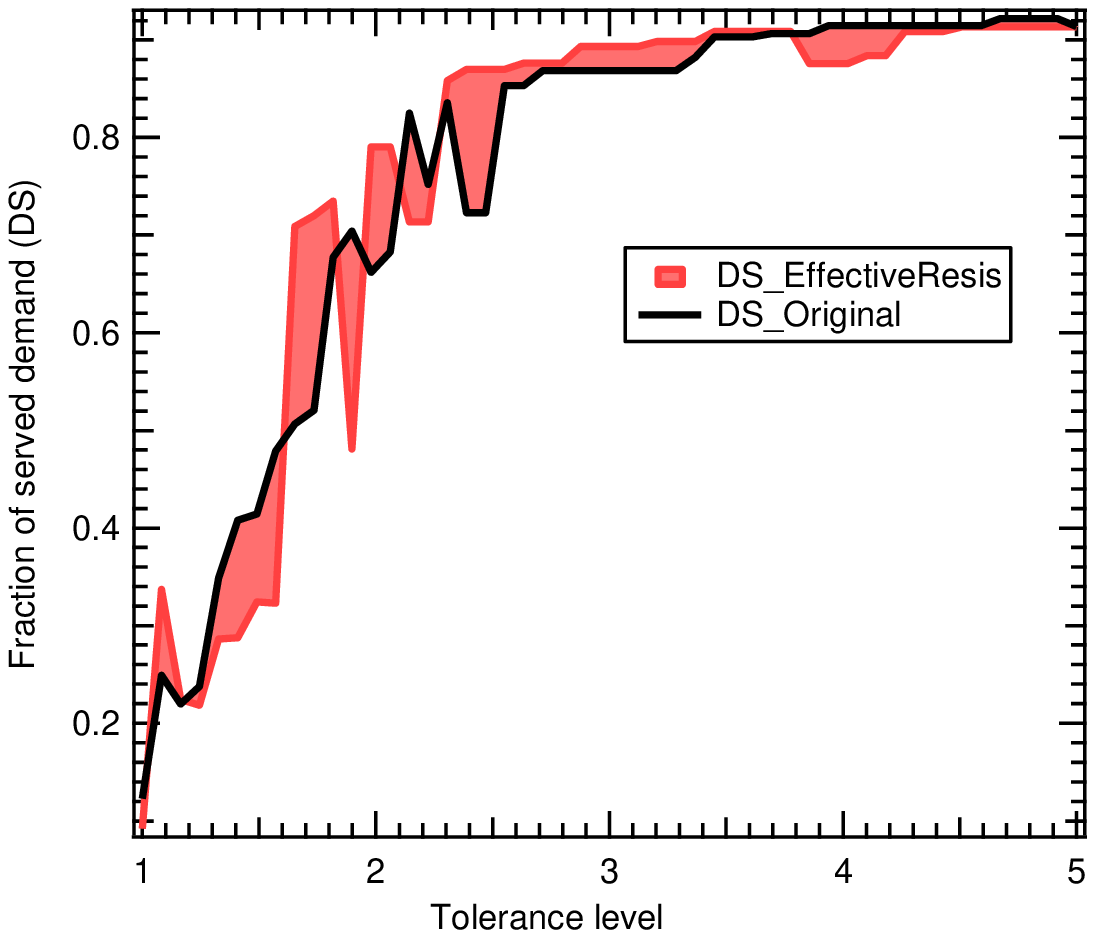}
\end{subfigure}
\begin{subfigure}[b]{0.491\columnwidth}
\includegraphics[width=\columnwidth]{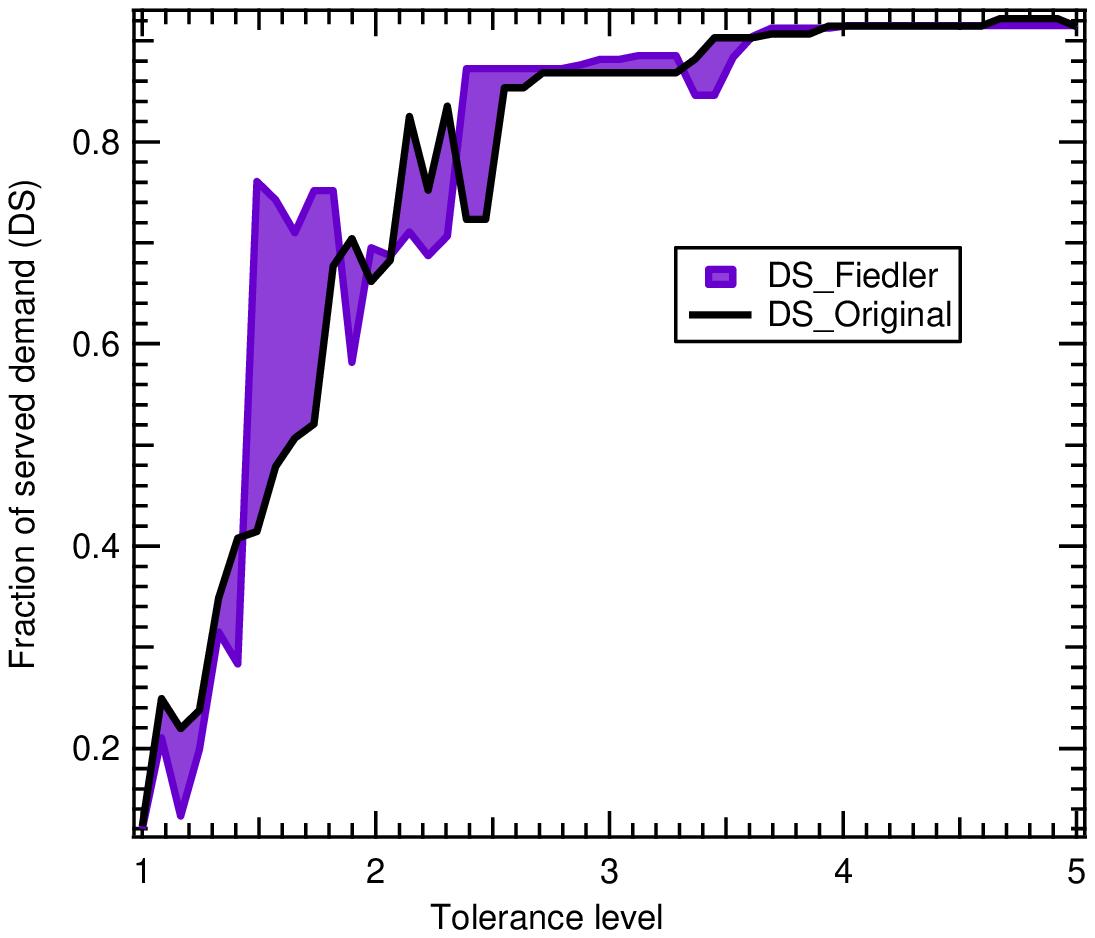}
\end{subfigure}
%\begin{subfigure}[b]{0.44\textwidth}
%\includegraphics[width=\textwidth]{Figures/Tolerance_DS_Optimal_EGR.eps}
%\end{subfigure}
%\begin{subfigure}[b]{0.44\textwidth}
%\includegraphics[width=\textwidth]{Figures/Tolerance_DS_ExhaustiveSearch.eps}
%\end{subfigure}
\caption{The performance of 
%the exhaustive search and 
the four strategies in IEEE 118 power system under different tolerance parameters. The attack strategy is based on the node significance centrality.}
\label{Tolerance_DS_NodeSigificanceAttack}
\end{figure}

\begin{figure}[!htp]\centering
\begin{subfigure}[b]{0.491\columnwidth}
\includegraphics[width=\columnwidth]{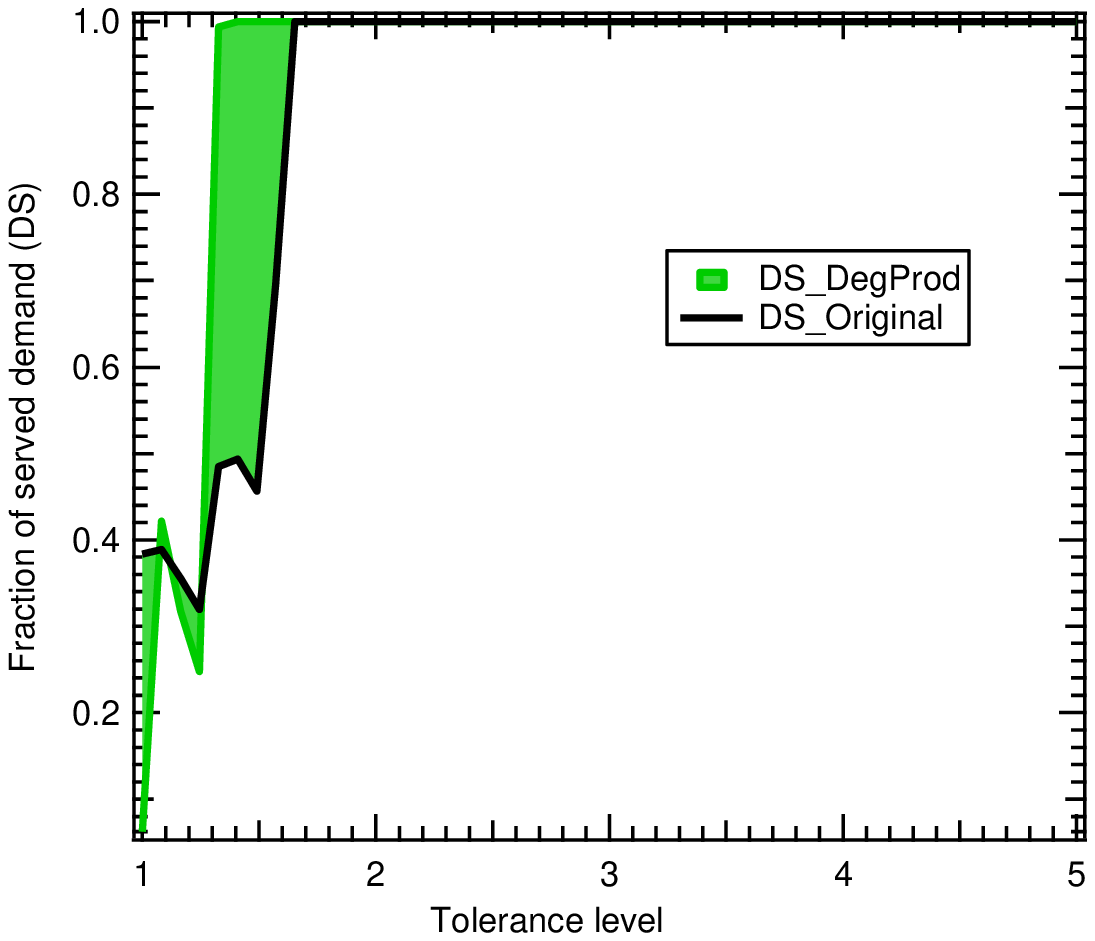}
\end{subfigure}
\begin{subfigure}[b]{0.491\columnwidth}
\includegraphics[width=\columnwidth]{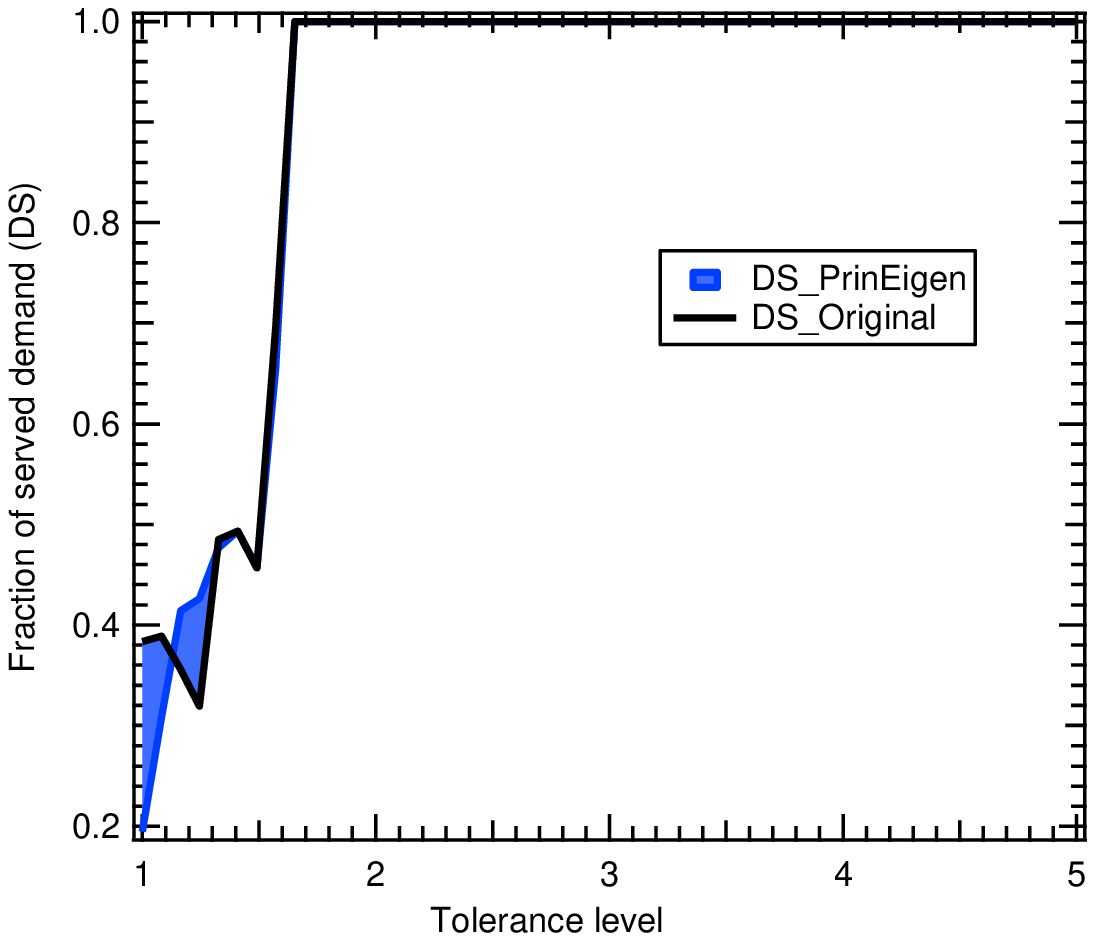}
\end{subfigure}
\begin{subfigure}[b]{0.491\columnwidth}
\includegraphics[width=\columnwidth]{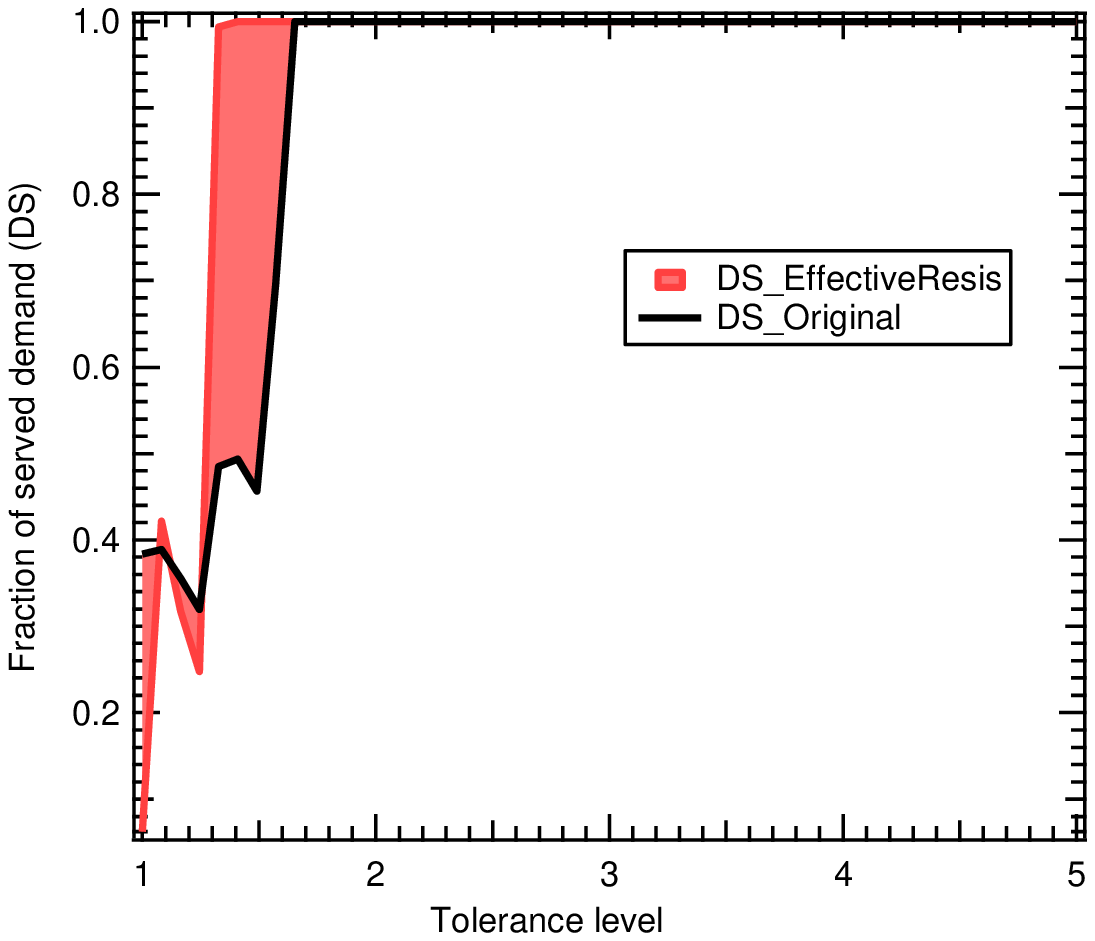}
\end{subfigure}
\begin{subfigure}[b]{0.491\columnwidth}
\includegraphics[width=\columnwidth]{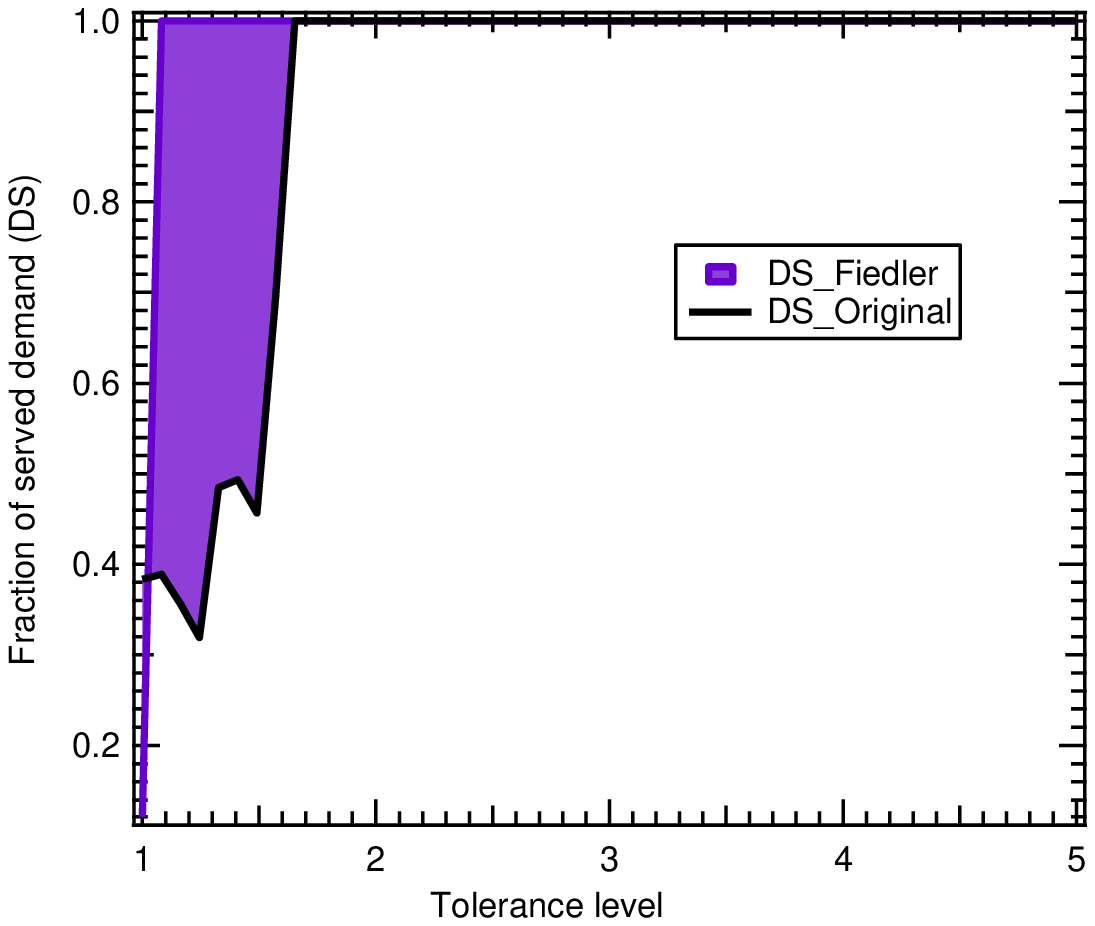}
\end{subfigure}
%\begin{subfigure}[b]{0.44\textwidth}
%\includegraphics[width=\textwidth]{Figures/Tolerance_DS_Optimal_BetweennessAttack.eps}
%\end{subfigure}
%\begin{subfigure}[b]{0.44\textwidth}
%\includegraphics[width=\textwidth]{Figures/Tolerance_DS_ExhaustiveSearch.eps}
%\end{subfigure}
\caption{The performance of 
%the exhaustive search and 
the four strategies in IEEE 118 power system under different tolerance parameters. The attack strategy is based on betweenness centrality.}
\label{Tolerance_DS_BetweennessAttack}
\end{figure}

To validate the results from Table \ref{CriticalAddedLine}, the original and improved IEEE 118 power systems are attacked based on the electrical node significance and the link betweenness, and damages after cascading failures are quantified. The improved power system refers to the system after adding a transmission line identified by strategies in Section \ref{Section_StrategiesAddingTransmissionLine}. Figures \ref{Tolerance_DS_NodeSigificanceAttack} and \ref{Tolerance_DS_BetweennessAttack} show the robustness curves for improved power grids under an interval of tolerance parameters $ [\alpha_{min}, \alpha_{max}] $ with $ \Delta_{\alpha}=0.05 $, and highlight the improvement of the grid robustness. In order to quantify the performance of the four strategies in improving the grid robustness, the robustness value $ r $ in equation \eqref{SimulationEvaluation} for each robustness curve is shown in Table \ref{Robustness_Value}. 

Figure \ref{Tolerance_DS_NodeSigificanceAttack} and Table \ref{Robustness_Value} show the performance of the strategies in the IEEE 118 power grid under the attack based on the node significance. The strategy based on the Fiedler vector has a robustness value $ r=0.777 $ which is an increase by $ 1.8\% $ compared to the original grid robustness (i.e. $ 0.763 $). The strategy based on the degree product and on the effective resistance have an equal performance. These two strategies have the same robustness value $ r =0.769 $ and increase the robustness by $ 0.8\% $. The strategy based on the principle eigenvector has the lowest performance and its robustness value is $ r=0.757 $ that decreases the robustness by $ 0.8\% $. 

Figure \ref{Tolerance_DS_BetweennessAttack} and Table \ref{Robustness_Value} present the performance of the strategies under the betweenness based attack. The strategy based on the Fiedler vector has the highest robustness value $ r=0.991 $, which is an increase by $ 8.2\% $ compared to the original grid robustness (i.e. $ 0.916 $). The strategy based on the degree product and on the effective resistance have an equal performance with the same robustness value $ r =0.949 $. The robustness is increased by $ 3.6\% $ compared to the original grid robustness. In contrast, the strategy based on the principle eigenvector with $ r=0.915 $ slightly decreases the robustness by $ 0.1 \% $. %In the scenarios of electrical node significance based attack and link betweenness based attack, the strategy based on the Fiedler vector has the highest performance in improving the power grid robustness. The strategy based on the degree product and on the effective resistance have a lower performance. In contrast, the strategy based on the principle eigenvector decreases the robustness rather than increases the robustness. 
The performance order of the strategies shown in Figures \ref{Tolerance_DS_NodeSigificanceAttack} and \ref{Tolerance_DS_BetweennessAttack} and Table \ref{Robustness_Value} is in agreement with the theoretical results in Table \ref{CriticalAddedLine}. 

\begin{table}[!htp]\centering
\begin{tabular}{cccc}
\hline \multirow{2}{*}{Strategy} & \multirow{2}{*}{line ID}& $ r  $ & $ r  $  \\
&  & (\text{Node Siginificance attack}) &(\text{Betweenness attack}) \\ 
\hline DegProd  & $ l_{87-117} $  & 0.769 & 0.949\\ 
  PrinEigen & $ l_{87-111} $  & 0.757 & 0.915 \\ 
  Fiedler & $ l_{111-117} $  & 0.777 & 0.991\\ 
  EffectiveResis & $ l_{87-117} $ & 0.769 & 0.949\\
%  Exhaustive Search & $ l_{12-105} $  & 0.792 & 0.951\\
\hline
\end{tabular} 
\caption{Critical lines identified by the four strategies 
%and exhaustive search 
and the robustness value $ r $ in IEEE 118 power system.}
\label{Robustness_Value} 
\end{table}

When the computational cost for finding the optimal links to add is prohibitive, the strategy based on the Fiedler vector with the highest performance is preferable compared to other strategies. Assuming that computing the Fiedler vector for large grids is not an option, the strategy based on the degree product can be an alternative. The degree based strategy is more likely to be chosen than the strategy based on the effective resistance due to the fact that these two strategies have comparable performance, while the strategy based on the degree product has lower computational complexity.
\subsection{Assessing the impact of the grid topology on Braess's paradox}
\label{subsection_BraessParadox}

Braess's paradox in this paper refers to the decrease of grid robustness by placing additional links. The relationship between the grid topology and the Braess's paradox in power grids is investigated.  

\begin{figure}[!htp]\centering
\includegraphics[width=0.32\columnwidth]{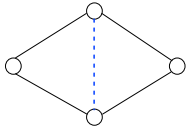}
\caption{Wheatstone bridge graph}
\label{WheatstoneStructure}
\end{figure}

The Wheatstone bridge graph (shown in Figure \ref{WheatstoneStructure}) refers to a graph consisting of four nodes, with four links creating a quadrilateral. A fifth link connects two opposite nodes in the quadrilateral, splitting the graph into two triangles \cite{calvert1993braess}. We consider the subgraph with four nodes and four links as the Wheatstone subgraph and the fifth link as the Wheatstone link. Braess's paradox indicates that the construction of the Wheatstone bridge graph by adding the Wheatstone link occasionally decreases the robustness of power grids. Let $ P_{\text{Wheatstone}} $ represent the percentage of the Wheatstone links and $ P_{\text{Paradox}} $ be the percentage of the links, whose addition results in Braess's paradox. In order to investigate the impact of the Wheatstone bridge graph on Braess's paradox, the correlation between the percentages $ P_{\text{Wheatstone}} $ and $ P_{\text{Paradox}} $ is quantified. The number of Wheatstone links is computed by the number of Wheatstone bridge subgraphs detected by FANMOD \cite{wernicke2006fanmod}, a tool for fast network motif detection.

\begin{figure}[!htp]\centering
\begin{subfigure}[b]{0.32\columnwidth}
\includegraphics[width=\columnwidth]{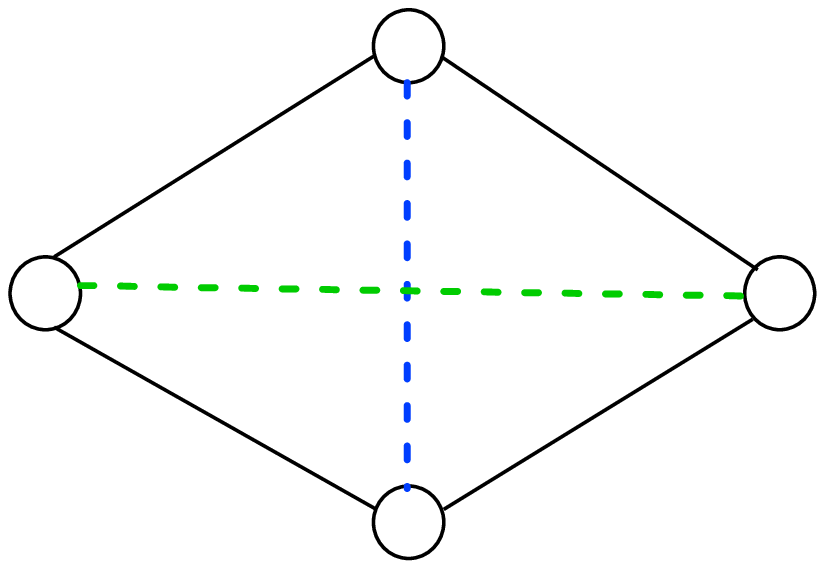}
\caption{Type I}
\end{subfigure}
\begin{subfigure}[b]{0.32\columnwidth}
\includegraphics[width=\columnwidth]{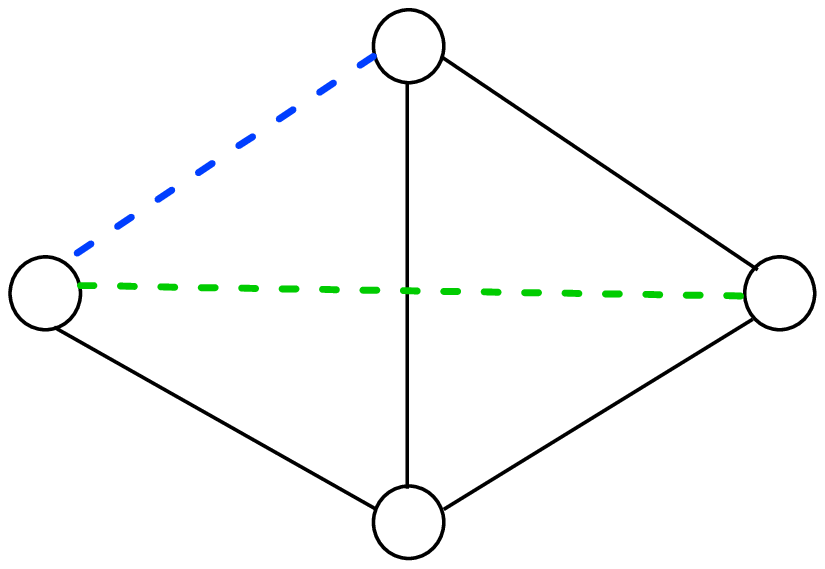}
\caption{Type II}
\end{subfigure}
\caption{Two types of subgraphs to build a Wheatstone bridge graph by adding the Wheatstone link. The dashed lines are the possible Wheatstone links.}
\label{WheatstoneTypes}
\end{figure}

Figure \ref{WheatstoneTypes} shows two types, Type I and Type II, of Wheatstone subgraphs from which a Wheatstone bridge graph is built by adding the Wheatstone link (the dashed line). For each subgraph, the number of the Wheatstone links is two times the total number of subgraphs of Type I and Type II. The percentage $ P_{\text{Wheatstone}} $ of Wheatstone links in all the possible added links $ L_c $ is computed by $ P_{\text{Wheatstone}}=\frac{2(N_{\text{TypeI}}+N_{\text{TypeII}})}{L_c} $, where $  N_{\text{Typek}} $ is the number of subgraphs of Type k. Table \ref{PercentageWheatstoneStructure} shows the percentage $ P_{\text{Wheatstone}} $ of Wheatstone links and the percentage $ P_{\text{Paradox}} $ in Figure \ref{Tolerance_DS_ExhaustiveSearch}. The correlation between $ P_{\text{Wheatstone}} $ and $ P_{\text{Paradox}} $ is $ 0.96 $ suggesting the criticality of the Wheatstone bridge graph (see Figure \ref{WheatstoneStructure}) to the occurrence of Braess's paradox.

\begin{figure}[!htp]\centering
\begin{subfigure}[b]{0.32\columnwidth}
\includegraphics[width=\columnwidth]{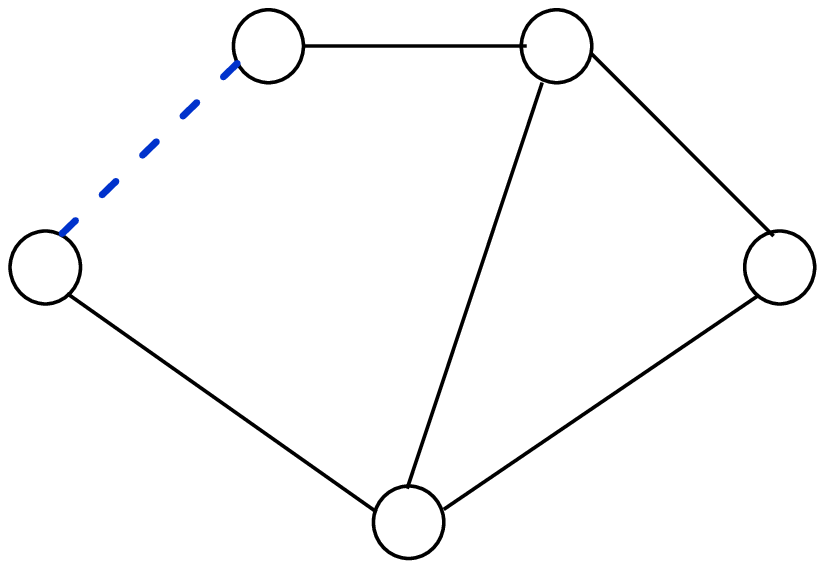}
\caption{Type III}
\end{subfigure}
\begin{subfigure}[b]{0.32\columnwidth}
\includegraphics[width=\columnwidth]{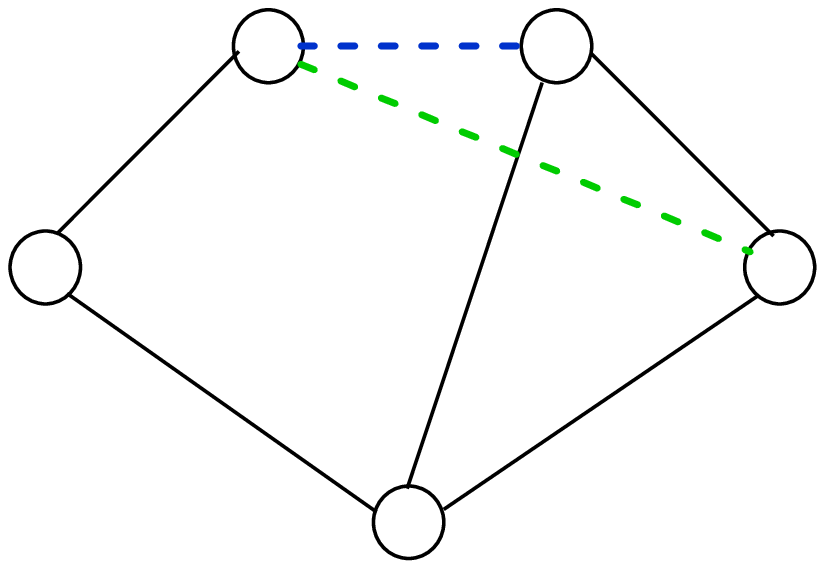}
\caption{Type IV}
\end{subfigure}
\begin{subfigure}[b]{0.32\columnwidth}
\includegraphics[width=\columnwidth]{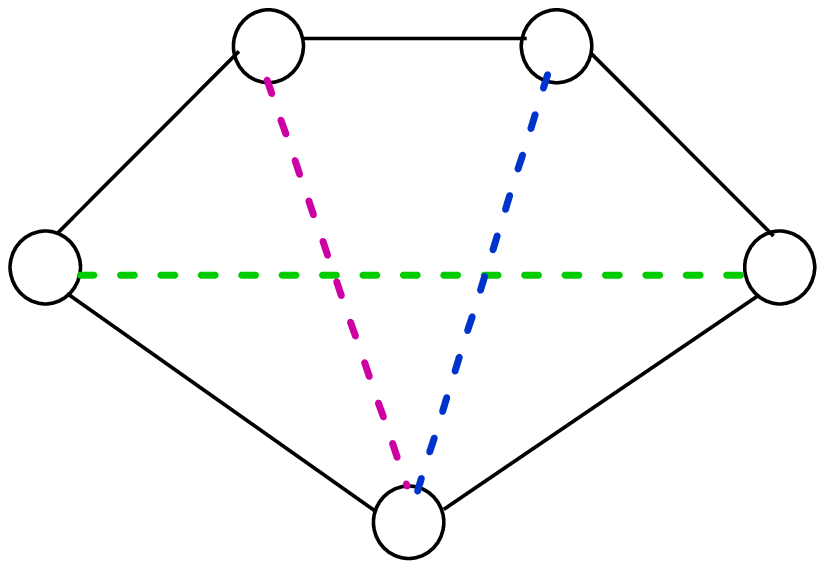}
\caption{Type V}
\end{subfigure}
\caption{Three types of subgraphs resulting in Braess's paradox by adding an extra link.}
\label{NewWheatstoneTypes}
\end{figure}

Besides the Wheatstone bridge graph that occasionally introduce Braess's paradox \cite{blumsack2007quantitative}, we further investigate other subgraphs that may lead to the Braess's paradox. Figure \ref{NewWheatstoneTypes} shows other three types, Type III to Type V, of subgraphs resulting in Braess's paradox when a single link is added. The dashed lines in Figure \ref{NewWheatstoneTypes} are the possible links that cause the Braess's paradox. Table \ref{PercentageNewStructure} shows the percentage $ P_{\text{Wheatstone}} $ after including the number of links added into Type III, IV and V. The percentage $ P_{\text{Wheatstone}} $ increases from $ 6.73\% $ to $ 25.00\% $ in IEEE 57 power system. An increase of the  $ P_{\text{Wheatstone}} $ from $ 4.53\% $ to $ 15.44\% $ is also observed in IEEE 118 and from $ 1.34\% $ to $ 4.11\% $ in IEEE 247 power system. Accordingly, the correlation between $ P_{\text{Wheatstone}} $ and $ P_{\text{Paradox}} $ increases to $ 0.971 $. The results indicate that the subgraphs from Type I to Type V provide an effective indication for the occurrence of the Braess's paradox in power grids. 
%The impact of other characteristics of power grids, such as the reactance values and the loading profiles, on the occurrence of Braess's paradox is still an open question. 

\begin{table}[!htp]\centering
\begin{tabular}{cccc}
\hline  & IEEE57 & IEEE118 & IEEE247 \\ 
\hline $ L_c $   & 1516    & 6717 & 30026\\
       $ N_{\text{TypeI}} $ & 0 & 20 & 30 \\ 
       $ N_{\text{TypeII}} $ & 51 & 132 & 171 \\  
       $P_{\text{Wheatstone}}$(\%) & 6.73 & 4.53  & 1.34  \\ 
       $ P_{\text{Paradox}}$ (\%)& 53.16 & 20.67 & 4.57 \\ 
\hline 
\end{tabular} 
\caption{The percentage $ P_{\text{Wheatstone}} $ and $ P_{\text{Paradox}} $ in IEEE power systems}
\label{PercentageWheatstoneStructure}
\end{table}
\begin{table}[!htp]\centering
\begin{tabular}{cccc}
\hline  & IEEE57 & IEEE118 & IEEE247 \\ 
%\hline $ L_c $   & 1516    & 6717 & 30026\\ 
 \hline      $ N_{\text{TypeIII}} $ & 95 & 256  & 299  \\ 
       $ N_{\text{TypeIV}} $ & 91 & 216 &  255 \\ 
       $ N_{\text{TypeV}} $ & 0 & 15 &  8 \\ 
       $P_{\text{Wheatstone}}$(\%) & 25.00 & 15.44  & 4.11  \\ 
%       $ P_{\text{Paradox}}$ (\%)& 53.16 & 20.67 & 4.57 \\ 
\hline 
\end{tabular}  
\caption{The percentage $ P_{\text{Wheatstone}} $ and $ P_{\text{Paradox}} $ in IEEE power systems}
\label{PercentageNewStructure}
\end{table}
\section{Conclusion and Discussion}
\label{Section_Conclusion}
This paper investigates the effective graph resistance as a metric for network expansions to improve the grid robustness against cascading failures. The effective graph resistance takes the multiple paths and their ability to accommodate power flows into account to quantify the robustness of power grids. The experimental verification on IEEE power systems demonstrates the effectiveness of the effective graph resistance to identify single links that improve the grid robustness against cascading failures. Additionally, when computational cost for finding optimal links is prohibitive, strategies that optimize the effective graph resistance can still identify an added link resulting in a higher level of robustness. Specifically, the strategy based on the Fiedler vector performs the best compared to other strategies and increases the robustness by $ 8.2\% $ in IEEE 118 power system under the betweenness based attack, while reduces the computational complexity from $ O(N^5) $ to $ O(N^3) $. 

The occurrence of Braess's paradox in power grids suggests that the robustness can be occasionally decreased by placing additional links. In particular, a badly designed power grid may cause enormous costs for new lines that actually reduce the grid robustness. The experimental results in this paper provide insights in designing robust power grids while avoiding the Braess's paradox in power grids.


\section*{Acknowledgment}

This research was supported by the China Scholarship Council (CSC) and by the NWO project RobuSmart, grant number 647.000.001.



%\bibliographystyle{IEEEtran}
%\begin{thebibliography}{1}
%\end{thebibliography}

\bibliographystyle{abbrv}
\bibliography{Bib/biblio}

\end{document}